# Implementation of Machine Learning Algorithms for Seismic Events Classification

Alemayehu Belay Kassa
*Electrical and Computer Engineering, Morgan State University, Baltimore, United States of America*
alkas2@morgan.edu

Mulugeta Tuji Dugda
*Electrical and Computer Engineering, Morgan State University, Baltimore, United States of America*
mulugeta.dugda@morgan.edu

*Abstract*

 The classification of seismic events has been crucial for monitoring underground nuclear explosions and unnatural seismic events as well as natural earthquakes. This research is an attempt to apply different machine learning (ML) algorithms to classify various types of seismic events into chemical explosions, collapses, nuclear explosions, damaging earthquakes, felt earthquakes, generic earthquakes and generic explosions for a dataset obtained from IRIS-DMC. One major objective of this research has been to identify some of the best ML algorithms for such seismic events classification. The ML algorithms we are implementing in this study include logistic regression, support vector machine (SVM), Naïve Bayes, random forest, K-nearest neighbors (KNN), decision trees, and linear discriminant analysis. Our implementation of the above ML classifier algorithms required to prepare and preprocess the dataset we obtained so that it will be fit for the ML training and testing applications we sought. After the implementation of the ML algorithms, we were able to classify the seismic event types into seven classes in the dataset, and a comparison of each classifier is made to identify the best algorithm for the seismic data classification. Finally, we made predictions of the different event types using the different classifier algorithms, and evaluated each of the various classifier algorithms for seismic prediction using different evaluation metrics. These evaluation metrics helped us to measure the performance of each algorithm. After implementing the seven ML algorithms and a comparison among those various ML algorithms, it has been demonstrated that the best accuracy among these classifiers happened for the Random Forest (RF) algorithm, with an accuracy of 93.5%.

## I. Introduction

Artificial Intelligence (AI) or Machine Learning (ML) has been playing an important role in the development of new approaches for seismic studies. Seismic studies include the study of earthquakes (natural and man-made shaking of the earth), seismic waves and what those waves tell us about the internal structure and composition of the Earth. Moreover, seismic signal processing has played an important role in the detection and monitoring of underground nuclear explosions. Both global and exploration seismology have been employing various ML algorithms over the last three decades at various levels [1- 4]. Exploration seismology in particular has been implementing all sorts of cutting-edge technologies including AI and ML [5-7] for the purpose of exploring the earth locally. In this study we will be making use of global seismology data analyzed at a regional level. Seismology is a data-driven science and its most important discoveries usually result from analysis of new data sets and/or development of new data analysis methods [8]. The classification of seismic events is crucial for monitoring underground nuclear explosions and detecting regional unnatural seismic events and also to classify tectonic earthquakes, explosions and mining induced earthquakes [9].

 Seismic signal processing has been a data rich science. Application of machine learning for gaining new insights from such rich seismic data processing is a rapidly evolving sub-field in seismology [10]. The main objective of this research is to apply different machine learning (ML) algorithms to classify seismic event types such as chemical explosions, collapses, nuclear explosions, damaging earthquakes, felt earthquakes, generic earthquakes and generic explosions in a dataset obtained from IRIS-DMC the data is called Earth scope.

 In the past few years, different seismic event classification studies using ML algorithms have been published. However, most of those papers have two or three classes, and few of them have four classes. Moreover, many of those published works use very few ML algorithms for classification. In this research we implemented seven machine learning algorithms and also we constructed seven classes, those classes being Chemical Explosions, Collapses, Damaging Earthquake, Felt Earthquake, Nuclear Explosions, Generic Earthquakes and Generic Explosions.

Classifying seismic events accurately and identifying seismic event types is highly critical, especially identifying those damaging earthquakes in a timely manner for the safety and wellbeing of societies and communities near and around earthquake prone areas. Such seismic event classification helps to set off an alarm whenever there is some kind of risk and helps for an early warning of people in those vulnerable areas. Moreover, it is crucial to distinguish between an underground nuclear explosion testing from other kinds of seismic events, to enforce the nuclear test ban treaty, which in turn might help to avoid such explosives from falling into bad and irresponsible actors.

This research mainly focuses on classifying seismic event types using different machine learning algorithms. First we compare different machine learning algorithms that are used here for such seismic classification. Then, we select the best algorithm to classify and predict the seismic events with good accuracy. After applying seven selected classical machine learning algorithms, comparison among those various classical machine learning algorithms have demonstrated that we can obtain the best accuracy using Random Forest (RFC), with an accuracy of 93.5%.

The paper is organized into seven sections. Section 2 talks about literature review, Section 3 discusses the methodologies applied in this paper. Section 4 describes the data used in this work and related analysis. Section 5 discusses the results and provides an analysis. Section 6 provides a discussion and future work. Finally, section 7 makes concluding remarks.

## II. Literature Review

This study is not intended to cover a comprehensive review of ML classification algorithms. However, in order for us to make a determination of which algorithms we might consider in this research and get the most benefit out of those, we made a short survey of literature which made some implementations of ML classification algorithms in the area of seismic studies. Classification is generally the process of predicting a class or category from some given data points. In other words, classification involves the process of predicting a certain class/category of events from observed values of features [11].

Over the last three decades, different seismic classification works employing different machine and deep learning algorithms have been published and we reviewed some of those research in this article, especially the most recent ones. The first paper we reviewed was written by Wang, T., Y. Bian, Y. Zhang, and X. Hou (2022) entitled "Using Artificial Intelligence Methods to Classify Different Seismic Events." This paper established three-class models to classify earthquakes, explosions, and mining-induced earthquakes using support vector machine (SVM) and long short-term memory network (LSTM) for classification.

The second paper we reviewed was written by Tang, L., Zhang, M., & Wen, L. (2020) entitled "Support vector machine classification of seismic events in the Tianshan orogenic belt." This paper established three class models to classify earthquakes, explosions, and mining induced earthquakes using support vector machines (SVM). The researchers in that article made use of seismic data from 32 stations for a total of 30,181 data points for classifications.

The third paper we reviewed was written by Sangkyeum Kim, Kyunghyun Lee and Kwanho You (2020) entitled "Seismic Discrimination between Earthquakes and Explosions Using Support Vector Machines (SVM)". This paper established two class models to classify earthquakes and explosions using support vector machines (SVM).

The fourth paper we reviewed was written by Linville, L., Pankow, K., & Draelos, T. (2019). "Deep learning models augment analyst decisions for event discrimination". That paper established two class models to classify quarry blasts and earthquakes using convolutional and recurrent neural networks.

The fifth paper we reviewed was written by Mousavi, S. M., & Beroza, G. C. (2020). "A machine-learning approach for earthquake magnitude estimation." The main goal of that paper is to develop a method for a fast and reliable estimation of earthquake magnitude directly from raw seismograms recorded on a single station. They use convolutional and recurrent neural networks for magnitude estimation. We have further reviewed some more related works and we summarized those as shown in Table 1.

| Ref. No. | Titles | Objectives | Methods | Results |
|---|---|---|---|---|
| [9] | Using Artificial Intelligence Methods to Classify Different Seismic Events | Uses five AI methods to classify seismic events such as earthquakes, explosions and mining-induced earthquakes | Use five AI methods including Support Vector Machine, Long Short Term Memory Network and others | Established three class models to classify earthquakes, explosions and mining-induced earthquakes |
| [13] | Seismic Data Classification using Machine Learning | Determine the time that P wave and S wave reach | Decision tree Random forest Support Vector Machine | Predicting the P and S wave arrival times. |
| [14] | Analysis of Seismic data using Machine Learning Algorithms | Comparing K-Nearest Neighbors (KNN) and Random Forest Algorithms for prediction. | K-Nearest Neighbors (KNN) Random Forest | Random Forest is better than K-Nearest Neighbors (KNN) for such a dataset. |
| [15] | Automatic classification of seismic events within a regional seismograph network | Classify earthquake and non-earthquake seismic events using Support Vector Machine (SVM) | Support Vector Machine (SVM) | Distinguishing natural earthquakes from human made seismic events. |
| [16] | Earthquake multi classification detection based velocity and displacement data filtering using machine learning algorithms | Earthquake multi classification detection using machine learning algorithms. | Support Vector Machine Random Forest (RF) Decision Tree (DT) Artificial Neural Network (ANN) | ANN algorithm is the best algorithm to distinguish between earthquake and non-earthquake, and vandalism vibrations |
| [20] | Support Vector Machine Classification of Seismic Events in the Tianshan orogenic belt | Develop a machine learning method using SVM method to classify tectonic earthquakes (TEs), quarry blasts (QBs) and induced earthquakes (IEs) | Support Vector Machine (SVM) | Established three class models to classify earthquakes, explosions and mining-induced earthquakes |
| [21] | Seismic Discrimination between Earthquakes and Explosions Using Support Vector Machine | To discriminate between earthquakes and explosions using Support Vector Machine (SVM) | Support Vector Machine (SVM) | Established two class models to classify earthquakes and explosions |
| [22] | Deep Learning Models Augment Analyst Decisions for Event Discrimination | To discrimination of quarry blasts and earthquakes. | Convolutional Neural Network Recurrent Neural Network | Established two class quarry blasts and earthquakes. |
| [23] | A Machine Learning Approach for Earthquake Magnitude Estimation | Develop a method for a fast and reliable earthquake magnitude estimation. | Convolutional Neural Network Long-Short Term-Memory (LSTM) | Earthquake magnitude estimation prediction. |

**Table 1:** Some Recent AI/ML Studies for Seismic Signals and Events Classification and Detection

### III. Methodologies

The identification of natural earthquakes and artificial blast events has been widely studied over the years. A variety of identification criteria, including body wave magnitude and surface wave magnitude, have been proposed and utilized over the years [18]. Moreover, a large body of research on earthquake classification implementing machine learning methods and making use of a large amount of seismic waveform data has been conducted in the field of seismology [18]. Classification in general is the process of predicting a certain class of events from observed values of features [18]. To implement a classification algorithm, first we need to train the classifier. For the purpose of this classification research, we have to divide our dataset into two parts: a training set and a test set. In this study, we used 70% of the dataset for training and 30% for testing. Our research implemented the machine learning (ML) algorithms listed and explained in the following subsections A-G. After discussing each of these ML algorithms, we will finally make a comparison between each of the algorithms and select the best ML algorithm based on performance.

### A. Logistic Regression

Logistic regression is a supervised learning classification algorithm that is used to predict the probability of a target variable [18]. Logistic Regression is a commonly employed classifier, used to assign the target value to a set of classes or categories. In this classifier, two types of classification have to be distinguished, which produces a binary classification, in which the output can take two classes, and the multiclass classification produces multi class outputs [18].

### B. Linear Discriminant Analysis

Linear Discriminant Analysis (LDA) is a very common technique for dimensionality reduction problems as a preprocessing step for machine learning and pattern classification applications. The LDA technique is developed to transform the features into a lower dimensional space, which maximizes the ratio of the between class variance to the within class variance, guaranteeing maximum class separability [26].

The goal of the LDA technique is to project the original data matrix onto a lower dimensional space. LDA technique is used to find a linear transformation that discriminates between different classes. Due to the high number of features or dimensionality, the LDA technique has been applied on biometrics, agriculture and medical applications [26].

### C. Support Vector Machine (SVM)

Support vector Machine (SVM) is a very powerful and diverse machine learning model, capable of performing linear, nonlinear, regression and classification [16]. SVM model is basically a representation of different classes in a hyperplane in multidimensional space. The goal of SVM is to divide the datasets into classes to find a maximum marginal hyperplane [11].

SVM is best used for binary classification problems. It is easier to visualize data in the higher dimension. To separate the data clearly into two classes a hyperplane is found. This hyperplane has a very special characteristic of Margin. This margin should be such that it is at a maximum distance from both sets of data points. The reason for that is it helps avoid overlapping of the two classes [13].

### D. Decision Trees

Decision tree analysis is a predictive modeling tool that can be applied across many areas. It can be constructed by an algorithmic approach that can split the dataset in different ways based on different conditions. Decisions trees are the most powerful algorithms that fall under the category of supervised algorithms [11]. It can be used for regression as well as classification problems [13].

This algorithm can also be used for data with multiple outputs. It performs data classification by forming a tree. Starting from the root node to the leaf node. At each node, there is information on the features that are used as conditions for determining the direction of data flow and the number of samples that arrive at the node, the class prediction value, and the class of the data at that node [16].

### E. Random Forest

Random forest is a supervised learning algorithm which is used for classification and for all types of problems [13]. A random forest algorithm creates decision trees on data samples and then gets the prediction from each of them and finally selects the best solution by means of voting. It overcomes the problem of overfitting by averaging. Random forests work better for a large range of data items than a single decision tree. Random forests are very flexible and possess very high accuracy [11].

### F. K-Nearest Neighbors (KNN)

K-nearest neighbors (KNN) algorithm is a type of supervised ML algorithm which can be used for classification. KNN algorithm uses feature similarity to predict the values of new data points which further means that the new data point will be assigned a value based on how closely it matches the points in the training set [11]. KNN is used for bigger datasets to determine the accuracy of the model [14].

### G. Naïve Bayes

The Naïve Bayes algorithm is a classification technique based on applying Bayes' theorem with a strong assumption that all the predictors are independent of each other. Naïve Bayes classifier having the assumption that the data from each label is drawn from a simple Gaussian distribution [11].

The Naïve Bayes model has three types of Gaussian Naïve Bayes. Gaussian Naïve Bayes, it is the simplest Naïve Bayes classifier having the assumption that the data from each label is drawn from a simple Gaussian distribution. Multinomial Naïve Bayes, another useful Naïve Bayes classifier is Multinomial Naïve Bayes in which the features are assumed to be drawn from a simple Multinomial distribution. Another important model is Bernoulli Naïve Bayes in which features are assumed to be binary (0s and 1s) [11].

## IV. Data Analysis

Data has been an essential component for any research. Getting high quality data for researchers, especially in the area of seismic studies and machine learning, is one of the biggest challenges. Using quality data in turn leads to a high quality and well received research. Seismology is a data rich and data driven science. Application of machine learning for gaining new insights from seismic data is a rapidly evolving approach in seismology. The availability of a large amount of seismic data and computational resources, together with the development of advanced techniques can foster more robust models and algorithms to process and analyze seismic signals [10]. However, some of the oldest seismic data have some challenges to access and to reuse for machine learning research.

We prepare and preprocess the seismic data we received from IRIS DMC, which was obtained by first requesting and then downloading from the IRIS DMC website. The dataset was prepared about 20 years ago and it includes several decades of nuclear explosion and regional seismic data. It has waveforms files, arrival table, event information table, event types table and other metadata information. The data and metadata are not uniform and placed on different tables and file formats. It should also be noted that the labels, headers, types and events category are not placed on single files and uniform format. We preprocessed and passed it through different stages to make the data uniform, reusable and easily applicable for this research. To prepare and preprocess the data, we have made use of machine learning data preprocess and preparing techniques.

To prepare the data, first it needs to understand the dataset. Understanding the data involves looking at the raw data, checking the dimensions of the data, and visualizing and summarizing the data statically. As mentioned, one way to understand the data is through visualization. Visualizations enable us to understand the distribution of attributes and their correlations within the data. Histogram plots, box and whisker plots and scatter plots help us to better visualize the data, the attributes and the correlations. After preprocessing and preparing the data, we finally use a csv file that has different attributes and classification categories.

The original file didn't have column names and it has some missing values. First, we give names for the columns based on the dictionaries of the dataset. Then, we removed those columns with mostly undefined values. Finally, we convert the files from other formats to csv format. The original file was in .origin format. Finally, our dataset has 13 columns and 460 rows.

### A. Dataset

As explained previously, the dataset was in .origin format originally and it didn't have column names and also it had some missing values. First, we assigned column names based on the dictionaries of the dataset made available by the original data authors. The data was 26 columns and 633 rows originally. Then, we removed those columns with omitted values. Finally, we converted the files from other formats to a csv format. As mentioned in the previous section, the dataset currently has 13 columns and 460 rows.

The current attributes of the dataset includes Latitude, Longitude, Depth, Time, OriginID, EventID, JulianDate, Nde, Gr, Sr, Mb (Body wave magnitude), Ml(Local magnitude) and Evtype. The dataset Body wave magnitude (Mb) has values in the range of 2.17 and 6.5. The local magnitude (Ml) on the other hand has a value range from 1.18 to 7.1.

The dataset has a minimum depth of 0.00 km and a maximum depth of 29.7 km. The Latitude ranges from 33.019 to 41.529. The Longitude ranges from -122.413 to -109.073. The class of the dataset is Evtype and we have a total of seven classes. Chemical Explosions has 60 events, Nuclear Explosions has 160 events, Generic explosion has 74 events, Collapses has 4 events, Damaging Earthquake has 6 events, Felt Earthquake has 17 events and Generic Earthquakes has 138 events. Here is the link for both unprocessed and preprocessed datasets: https://github.com/Alemayehubk/Seismic-Events-Data.

Figure 1 illustrates the attributes distribution of the dataset using histogram plots, and each graph shows each of the attribute distributions as well as the types of distributions.

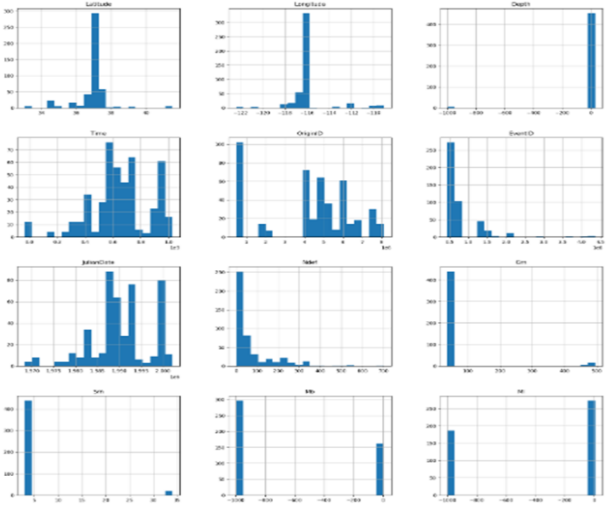

**Figure 1:** The attributes distribution of the dataset using histogram plots, from the graph it shows each attribute distribution and types of distributions.

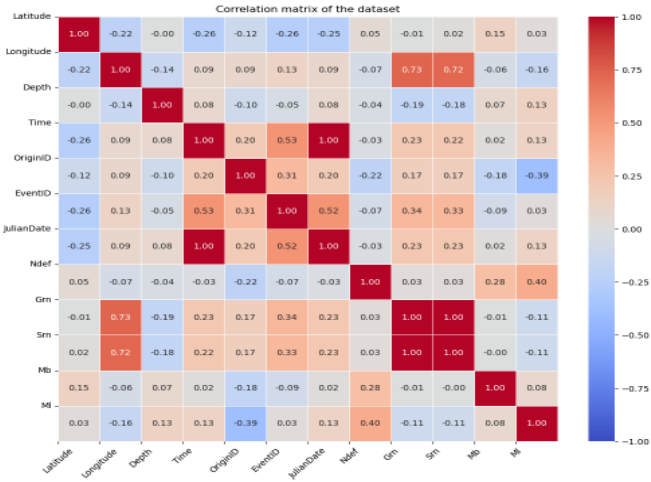

**Figure 2:** The Correlation matrix of the dataset

Figure 2 displays the correlation matrix of the dataset. Correlation is an indication about the changes between two variables [11]. From the above correlation matrix in Figure 2, it can be inferred that the bottom left is symmetrical with the top right and correlated with each other. It shows positive correlation with each other's attributes.

## V. Results and Analysis

In this paper, we applied machine learning algorithms to classify seismic events. We establish seven classes: Chemical Explosions, Nuclear Explosions, Generic explosion, Collapses, Damaging Earthquake, Felt Earthquake and Generic Earthquakes. These are seismic events in a dataset obtained from IRIS-DMC, currently known as Earth scope. The algorithms of Machine learning we are using such as Logistic Regression, Linear Discrimination, Support Vector Machine (SVM), Naïve Bayes, Decision tree, K-Nearest Neighbors and Random Forest algorithms. We divide the dataset into two parts: a training dataset and test dataset. We take 70% of the data for training purposes and 30% for testing purposes.

There are various metrics which we have used to evaluate the performance of the machine learning algorithms classification [11]. To compare the performance of each algorithm, to select the best algorithm we use Accuracy, Precession, Recall and F1-Score to evaluate each algorithm.

**Accuracy:** It is the most common performance metric for classification algorithms. It is defined as the number of correct predictions made as a ratio of all predictions made [11].

$$Accuracy = (TP + TN)/(TP + FP + FN + TN)$$

Where TP is a True Positive, TN is a True Negative, FN is a False Negative and FP is False Positive.

**Precision:** It is the number of correct documents returned by the model.

$$Precision = TP/(TP + FP)$$

Where TP is a True Positive and FP is False Positive.

**Recall:** It is the number of positives returned by the model.

$$Recall = TP/(TP + FN)$$

Where TP is a True Positive and FP is a False Negative.

**F1 Score:** This score will give us the harmonic mean of precision and recall. The best value of F1 would be 1 and worst would be 0 [11].

$$F1 = 2 * (Precision * Recall)/(Precision + Recall)$$

**Support:** Support is the number of samples of the true response that lies in each class of the target values [11].

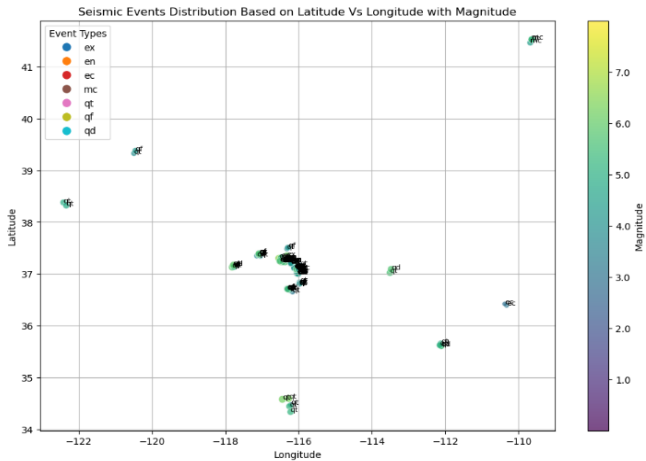

**Figure 3:** Seismic events distribution based on latitude and longitude. Here are the detailed event types which are shown with abbreviations in the above Figure: *ex* stands for generic explosion, *en* stands for nuclear explosion, *ec* stands for chemical explosion, *qt* stands for generic earthquake/tectonic, *qf* stands for felt earthquake, *mc* stands for collapse and *qd* stands for damaging earthquake.

### A. Machine Learning Algorithms

To compare the performance of each machine learning algorithm, to select the best algorithm. After comparison to classify and to predict the seismic events types. We use the evaluate metric to compare each algorithm. There are various metrics which we have used to evaluate the performance of the ML algorithms classification [11]. We use Accuracy, Precession, Recall and F1-Score to evaluate each algorithm. The following graph shows the comparison of different machine learning algorithms.

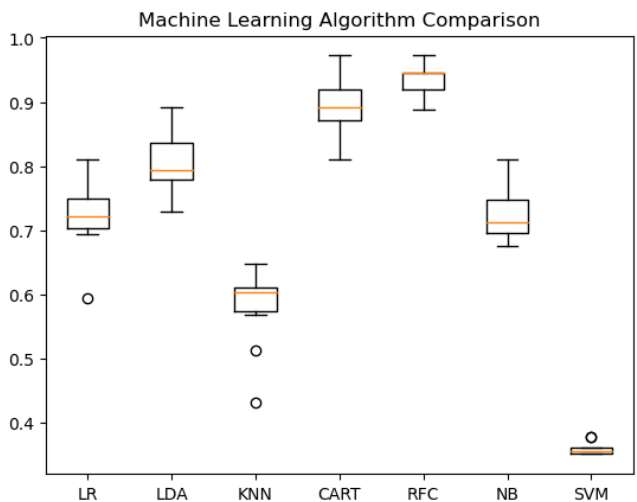

**Figure 4:** Comparison of each machine learning algorithm.

To make a comparison of accuracy results, with the implementation of Support Vector Machines (SVM), we have got an accuracy of 34.7%. Similarly, with K-Nearest Neighbors (KNN) we have got an accuracy of 59.4%, with Gaussian (NB) we have got an accuracy of 70.5%, with Logistic Regression (LR) we have got an accuracy of 71.5%, with Linear discriminant we have got an accuracy of 81%, with Decision Trees (CART) we have got an accuracy of 90%, and finally with Random Forest (RFC) we have got an accuracy of 93.5%. Therefore, from this comparison the best accuracy we have got comes from the Random Forest algorithm. The following table shows the classification report of the Random Forest Algorithm.

| Event Types | Precision | Recall | F1-score | Support |
|---|---|---|---|---|
| **Chemical Explosions** | 0.86 | 1.00 | 0.93 | 19 |
| **Nuclear Explosions** | 1.00 | 1.00 | 1.00 | 50 |
| **Generic Explosions** | 0.94 | 1.00 | 0.97 | 15 |
| **Collapses** | 1.00 | 0.50 | 0.67 | 2 |
| **Damaging Earthquake** | 0.50 | 0.50 | 0.50 | 2 |
| **Felt Earthquake** | 0.50 | 0.40 | 0.44 | 5 |
| **Generic Earthquakes** | 0.95 | 0.91 | 0.93 | 45 |

**Table 2:** Random Forest Algorithm Classification Report

We established seven classes for this particular study such as chemical explosions, nuclear explosions, generic explosions, collapses, damaging earthquakes, felt earthquakes and generic earthquakes.

Table 2 above is prepared specifically for a classification report implementing the random forest ML algorithm. The number of sample values for the support of Chemical Explosions is 19, which has a precision of 100%, a recall of 100% and an F1-Score of 100%. The number of sample values for the support of Nuclear Explosion is 50, which has a precision of 86%, a recall of 100% and an F1-Score of 93%. The number of sample values for the support of Generic Explosions is 15, which introduced 94% precision, a recall of 100% and an F1-Score of 97%. The number of sample values for the support of Collapse is 2, which has 100% precision, a recall of 50% and an F1-Score of 67%. The number of sample values for the support of Damaging Earthquake 2, which produced 50% precision, a recall of 50% and an F1-Score of 50%. The number of sample values for the support of Felt Earthquake is 5 with 50% precision, 40% recall and 44% F1-Score. The number of sample values for the support of Generic Earthquake is 45 with 95% precision, 91% recall and 93% F1-Score.

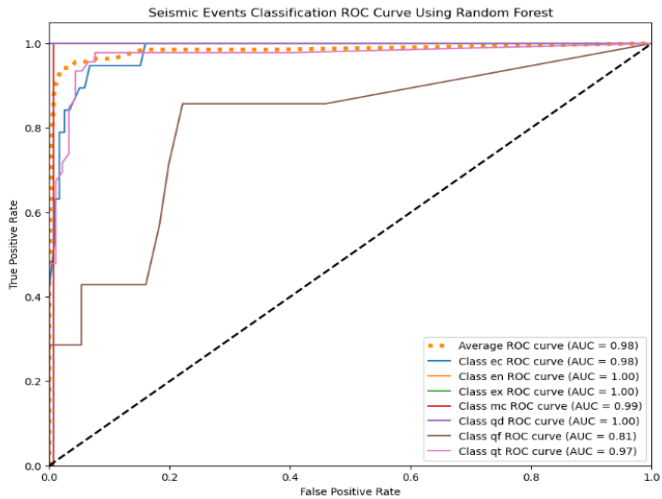

**Figure 5:** Seismic events classification ROC curve for Random Forest

The receiver operating characteristic curve (ROC) and the area under the curve (AUC) were adopted as effective measures of discrimination accuracy. The ROC curve is a graphical analysis tool that detects the threshold of the optimum model. Figure 5 displays the ROC for a random forest machine learning implementation. The process of plotting the ROC curve is to calculate the true positive rate (TPR) and the false positive rate (FPR) as the threshold of the classifier is changed. The TPR is equal to the ratio of the true positive (TP) correctly classified by the system to the sum of the TP and the false negative (FN) incorrectly classified by the system. TPR is obtained as TPR = TP/(TP + FN). FPR is defined as the ratio of the false positive (FP) to the sum of the FP and the true negative (TN), FPR = FP/(FP + TN). In the Roc curve, the AUC can be used to evaluate the performance of the classifiers [21]. In machine learning the larger the AUC value is the better the classifier [11].

As shown on Figure 5 above, all AUC values are greater than 0.9, which shows that the resulting prediction is very good. Based on the ROC curve result each event has the following AUC values: chemical explosions has 0.98, nuclear explosions has 1.00, generic explosions has 1.00, collapses has 0.99, damaging earthquake has 1.00, felt earthquake has 0.81 and generic earthquake has 0.97 AUC values.

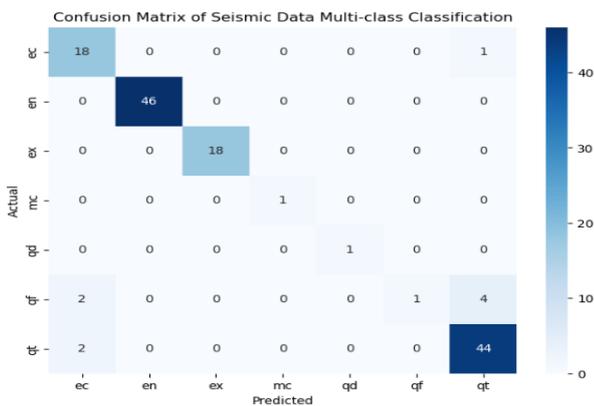

**Figure 6:** Confusion Matrix for multi-classification of seismic dataset

A confusion matrix is used to measure the performance of a classification problem where the output can be of any one type of the classes. A confusion matrix is a table with two dimensions, Actual versus Predicted, and furthermore, both dimensions have True Positives (TP), True Negatives (TN), False Positives (FP), False Negatives (FN) [11].

As shown in Figure 6 above, the confusion matrix shown for Chemical Explosions (**ec**) class, for instance, takes 18 values randomly for classification, out of which it has 0 false negative and 0 false positive. For Nuclear Explosion (**en**) it takes 46 values randomly for classification, out of which it has 0 false negative and 0 false positive. For Generic Explosions (**ex**), it takes 18 values for classification, out of which it has 0 false negative and 0 false positive. For Generic Earthquake (**qt**) it takes 44 values randomly, out of which it has 2 false positive and 1 false negative.

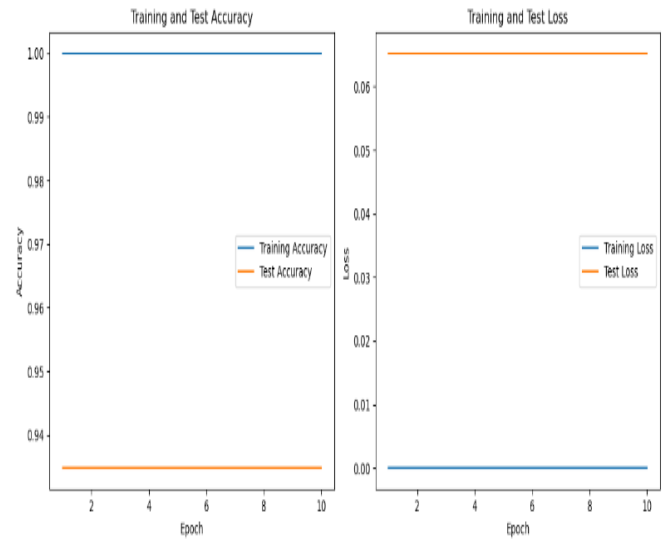

**Figure 7:** Seismic data multi class classification accuracy and loss graph.

To determine the **multi class classification accuracy** for our seismic data, we just have to determine which predictions were correct and which were not. Accuracy in fact is "blind" to specific classes. To calculate the accuracy, we need to **divide the number of all the correct predictions by the total number of predictions**. In this study, the accuracy is found to be 93.5% for random forest implementation.

On Figure 7**,** the graph on the left shows training and test accuracy, for which the training accuracy is 100% and the test accuracy is 93.5%. The graph on the right shows training and test loss, for which the training loss is 0 and the test loss is a very small number close to zero. Both training and testing accuracy and loss shows a linear graph and results.

## VI. Discussion and Future Works

### A. Discussion

The whole dataset which was obtained from the IRIS DMC was originally prepared and organized about 20 years ago. The dataset covers several decades of nuclear explosion and regional seismic data. It has waveform files, arrival table, event information table, event types table and other metadata information.

The final dataset we used for this specific research is a csv file and in the previous formats the data didn't have column names and it has some missing values. First we give column names based on the dictionaries of the dataset. Then, we removed the omitted columns and some missing values. Finally, we converted the files from other formats to csv format. The file was .origin format. Currently, the file has 13 columns and 460 rows. We develop models of machine learning algorithms and we made a comparison of performances. We discuss the result below from the lowest to the highest result we obtained using the following seven machine learning classifier algorithms.

Support Vector Machine (SVM) is a very powerful and diverse machine learning model, capable of performing linear, nonlinear, regression and classification [16]. Using Support Vector Machines (SVM) we have got an accuracy of 34.7%. SVM provided the lowest result for this particular research. In general, SVM is best suited for binary classification problems [11]. It is important to note that this study has multiclass classification and seven seismic event types or classes.

K-nearest neighbors (KNN) algorithm is a type of supervised ML algorithm which can be used for classification. K-nearest neighbors (KNN) algorithm uses feature similarity to predict the values of new data points which further means that the new data point will be assigned a value based on how closely it matches the points in the training set [11]. Using K-Nearest Neighbors (KNN) we have got an accuracy of 59.4%, which is generally used for bigger datasets to determine the accuracy of the model [14], while in our particular research work here we only have 13 columns and 460 rows for the dataset.

The Naïve Bayes (NB) algorithm is a classification technique based on applying Bayes' theorem with a strong assumption that all the predictors are independent of each other. Naïve Bayes classifier has the assumption that the data for each label is drawn from a simple Gaussian distribution [11]. For an implementation of Gaussian (NB) we have got an accuracy of 70.5%.

Logistic regression is a supervised learning classification algorithm used to predict the probability of a target variable [11]. It has two types of classification to be distinguished using logistic regression: the binary classification, in which the output can only take two classes, and the multiclass classification [25]. Our study is a multiclass classification and implementing Logistic Regression (LR) we have got an accuracy of 71.5%.

Linear Discriminant Analysis (LDA) is a very common technique for dimensionality reduction problems as a preprocessing step for machine learning and pattern classification applications [26]. Using linear discriminant we have got an accuracy of 81%.

Decisions trees are the most powerful algorithms that fall under the category of supervised algorithms [11]. It can be used for regression as well as classification problems [13]. It can also be used for data with multiple outputs [16]. Our current study involves multiclass classification and seven seismic event types of classes. We have got a very good result implementing Decision Trees (CART) with an accuracy of 90%.

Random forest is a supervised learning algorithm which is also used for classification and for all other types of problems [13]. It overcomes the problem of overfitting by averaging. Random forests work better for a large range of data items than a single decision tree. Random forests are very flexible and possess very high accuracy [11]. Using Random Forest (RFC) we have got an accuracy of 93.5%. Therefore, based on the comparison of the accuracies for the different machine learning algorithms, the best accuracy we have obtained for this particular study is when implementing the Random Forest algorithm.

### B. Future Works

The dataset used in this research was obtained from IRIS-DMC, currently known as Earth scope. The dataset is a csv file and it has 13 columns, 460 rows and seven classes. It has the following seven classes with the specified number of data points: Chemical Explosions has 60, Nuclear Explosions has 160, Generic explosion has 74, Collapses has 4, Damaging Earthquake has 6, Felt Earthquake has 17 and Generic Earthquakes has 138.

In the future, we will continue our study with a larger size of dataset that has those given seven seismic events and datasets with more data points. We will also implement deep learning and image processing algorithms to classify such seismic datasets. In addition, we will attempt to improve the classification accuracy and improve the calculation speed.

## VII. Conclusions

In this research we applied different classical machine learning (ML) algorithms to classify the seismic event types into chemical explosions, collapses, nuclear explosions, damaging earthquakes, felt earthquakes, generic earthquakes and generic explosions for a dataset obtained from IRIS-DMC, currently known as Earth scope. We have two major objectives in this study: the first objective is to identify the best algorithm for such seismic data classification among a selected set of ML models. The second objective is to make a classification with good accuracy based on the identified best ML algorithms.

Using the given dataset and the set of ML algorithms, we have obtained the best accuracy when implementing the Random Forest classifier and it has an accuracy of 93.5%. Using random forests, we could classify and predict those seven seismic events. From our experiment Random forest has the best performance and has the best accuracy for classification and prediction for such seismic events.

In the future, we will continue our study with a larger size of dataset that has those seven and even more seismic event classes with the best accuracy and performance.

## Data and Resources

The data used in this study was prepared about 20 years ago and it covers several decades of nuclear explosion and regional seismic data. The dataset was obtained from IRIS-DMC, currently known as Earth Scope.

## Acknowledgments

We would like to thank the Department of Computer Science (CS) and the Department of Electrical and Computer Engineering at Morgan State University, esp. the Masters in Advanced Computing Program at CS, and all the faculty in the CS Department. We are also grateful to the Computational Science Initiative (CSI) division at Brookhaven National Laboratory (BNL) in the USA for their collaborative research efforts, especially Dr. Line Pouchard and Dr. Yuewei Lin. We would also like to say thank you to Bill Walter and Michael E. Pasyanos of Lawrence Livermore National Laboratory (LLNL) for their support and encouragement.

We would also like to acknowledge the US National Science Foundation (NSF) for supporting this and other related research through a grant (2101080) awarded to Dr. Mulugeta Dugda. However, NSF is not responsible for any of the results published in this paper.

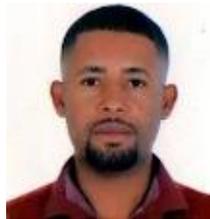

**Alemayehu Belay Kassa** received a BSc degree in Computer Science from Adama Science and Technology University, Ethiopia. He received an MSc degree in Software Engineering from HiLCoE, School of Computer Science and Technology, Ethiopia. He received an MS degree in Advanced Computing from Morgan State University, USA. Currently, he is a PhD Student and Graduate Research Assistant in Electrical and Computer Engineering at Morgan State University.

His research interests includes application of  Machine Learning, Deep Learning, Image Processing and Computer Vision on Seismic signal processing, Cyber Security and Embedded Systems.

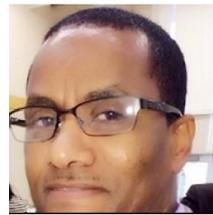

**Dr. Mulugeta Tuji Dugda** received BSc and MSc degrees in Electrical Engineering from Addis Ababa University, Ethiopia.  He earned PhDs in seismology (Pennsylvania State University) and electrical engineering (North Carolina A&T State University), USA.

Currently, he is a lecturer at Morgan State University, and his research interests include Seismic signal processing, Digital Signal Processing, Data Analytics, Big Data, Machine Learning (ML), Optimization and Communications, Cyber Security and Engineering Education.